\documentclass[aps,prl,twocolumn,showpacs,floatfix,superscriptaddress]{revtex4-1}
\usepackage{graphicx,color}
\usepackage{amsfonts}
\usepackage[figuresright]{rotating}
\usepackage{amssymb}
\usepackage{amsmath}
\usepackage{psfrag}
\usepackage{subfigure}
\usepackage{multirow}
\usepackage{tabularx}
\usepackage{textcomp}
\usepackage{units}
\usepackage{hyperref}
\hypersetup{
 pdfnewwindow=true, colorlinks=true,
 linkcolor=blue, anchorcolor=blue,
 citecolor=blue, filecolor=blue,
 menucolor=blue, urlcolor=blue}

\usepackage{graphicx}
\usepackage{dcolumn}
\usepackage{bm}
\usepackage{color}

\makeatletter

\begin{document}


\title{Quantum paraelastic two-dimensional materials}

\author{Tyler B.\ \surname{Bishop}}
\affiliation{Department of Physics, University of Arkansas, Fayetteville, AR 72701, USA}
\author{Erin E.\ \surname{Farmer}}
\affiliation{Department of Physics, University of Arkansas, Fayetteville, AR 72701, USA}
\author{Afsana\ \surname{Sharmin}}
\affiliation{Department of Physics, University of Arkansas, Fayetteville, AR 72701, USA}
\author{Alejandro\ \surname{Pacheco-Sanjuan}}
\affiliation{Departamento de Ingenier{\'\i}a Mec\'anica, Universidad T\'ecnica Federico Santa Mar{\'\i}a, Valpara{\'\i}so, Chile}
\author{Pierre\ \surname{Darancet}}
\affiliation{Center for Nanoscale Materials, Argonne National Laboratory, Argonne, IL 60439, USA}
\author{Salvador\ \surname{Barraza-Lopez}}
\email{sbarraza@uark.edu}
\affiliation{Department of Physics, University of Arkansas, Fayetteville, AR 72701, USA}
\affiliation{Center for Nanoscale Materials, Argonne National Laboratory, Argonne, IL 60439, USA}

\begin{abstract}
We study the elastic energy landscape of two-dimensional tin oxide (SnO) monolayers and demonstrate a transition temperature of $T_c=8.5\pm 1.8$ K using {\em ab-initio} molecular dynamics (MD), that is close to the value of the elastic energy barrier $J$ derived from $T=0$ K density functional theory calculations. The power spectra of the velocity autocorrelation throughout the MD evolution permits identifying soft phonon modes likely responsible for the structural transformation. The mean atomic displacements obtained from a Bose-Einstein occupation of the phonon modes suggest the existence of a quantum paraelastic phase that could be tuned with charge doping: SnO monolayers could be 2D quantum paraelastic materials with a charge-tunable quantum phase transition.
\end{abstract}

\date{\today}

\maketitle

Structural transformations in two-dimensional (2D) materials were first observed in transition metal dichalcogenides \cite{tmdm1,tmd0,tmd1} exposed to large gate and bias voltages \cite{tmdm1,tmd0,tmdelectrostatic}. These phase transitions induce two-dimensional ferroic and multiferroic behavior~\cite{Mehboudi2016,Seixas2016,Kai}. Three conditions underpin ferroic behavior: (i) the presence of structural degeneracies leading to a high-to-low-symmetry structural configuration as temperature ($T$) is decreased, (ii) a low-energy barrier $J$ along the path joining the degenerate ground states, and (iii) the existence of macroscopic monodomains \cite{Review}. The existence of macroscopic domains may not be granted when $J$ is too small.

With a predicted multiferroic behavior~\cite{Seixas2016}, SnO monolayers offer an ideal testbed for this observation:
Similar to 2D iron sulfide (FeS), SnO monolayers have a rectangular unit cell in their low-temperature ground state.
Neglecting quantum fluctuations, a transition temperature $T_c$ can be determined via molecular dynamics, beyond which these compounds turn into a traditional litharge structure; i.e., a square unit cell that otherwise maintains its atomistic coordination. But while the material behavior of FeS is driven by at least three competing energy scales of similar magnitude (the superconducting gap, spin-splitting/magnetic anisotropy barrier, and the elastic barrier leading to the rectangular-to-square unit cell transformation), the structural transformation of a pristine charge-neutral SnO monolayer can be characterized without concern for these competing energy scales.

In this Letter we study the ferro- to para-elastic transition in SnO monolayers, starting from the parametrization of their elastic energy landscape, a discussion of their structural degeneracies, and of their elastic energy barrier $J$. Using {\it ab initio} {calculations of the unit cell}, {\it ab initio} molecular dynamics (MD) calculations on the NPT ensemble, {and} model calculations, we demonstrate that the structural transition temperature $T_c$ is very similar to $J$.  The power spectra of the MD evolution permits identifying soft phonon modes likely responsible for the observed structural transformation. The Bose-Einstein occupations of the phonon modes lead to increased atomistic fluctuations that may stabilize a quantum paraelastic phase at low temperature that is tunable via electrostatic gating.

\begin{figure}[tb]
\begin{center}
\includegraphics[width=0.48\textwidth]{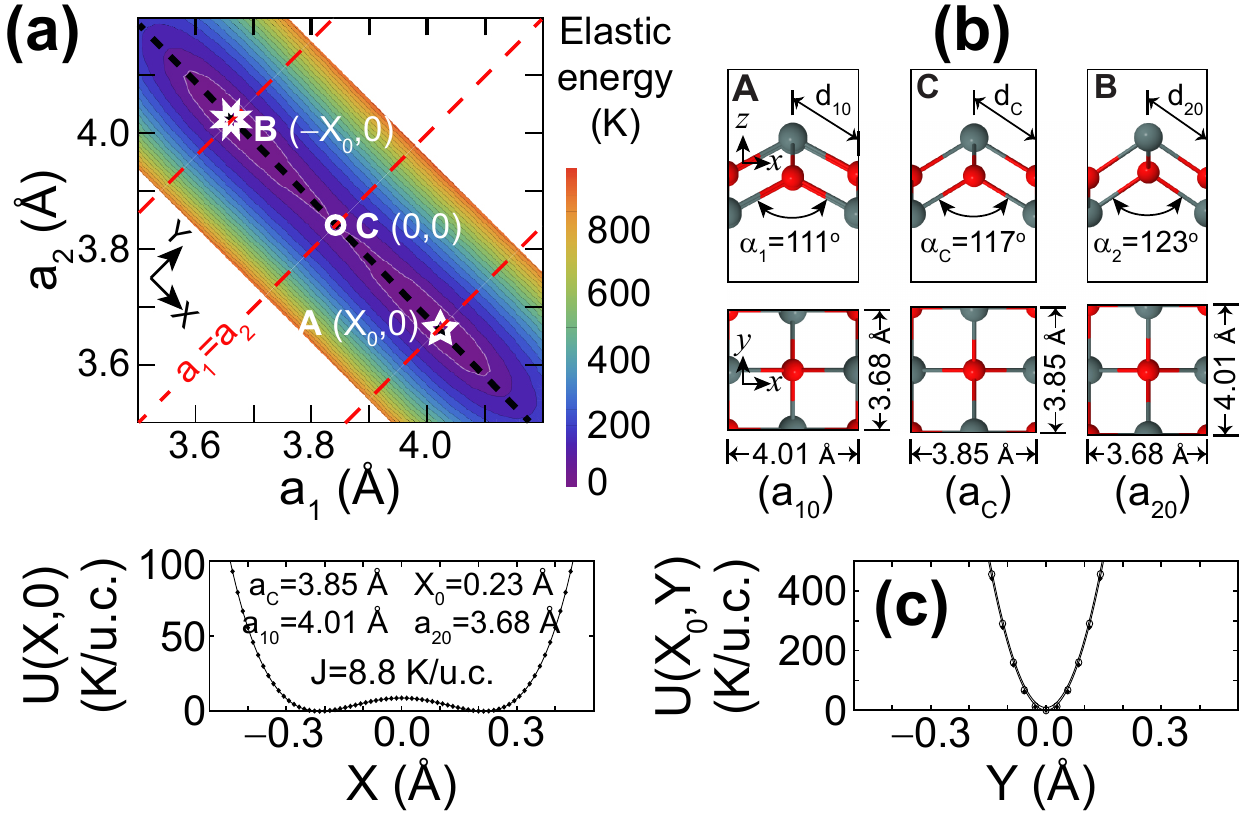}
\caption{(a) Elastic energy landscape for the SnO monolayer with zero doping. (b) Unit cells at the energy minimum (structures $A$ and $B$) and for the square structure at point $C$. Structural order parameters are shown. (c) Energy cuts through the black and red dashed lines shown in subplot (a).}\label{fig:figure1}
\end{center}
\end{figure}

Figure \ref{fig:figure1}(a), obtained with the { {\em VASP}} code \cite{vasp,vasppseudos} that implements the plane-wave pseudopotential approach within DFT with PBE exchange-correlation pseudopotentials \cite{PBE,vasppseudos}, shows the energy landscape for a SnO monolayer. {Optimizations were performed with forces smaller than $10^{-3}$ eV/\AA{} and using up to 90$\times 90\times 1$ k-points.} There, structure $A$ has a rectangular unit cell with lattice constants $a_{10}=4.01$ \AA{} and $a_{20}=3.68$ \AA, as highlighted by a star. O atoms have coordinates $\mathbf{b}_1$=(0,0,0) and $\mathbf{b}_2$=($a_{10}/2$,$a_{20}/2$,$-\delta_z$), while Sn atoms are located at $\mathbf{b}_3$=($a_{10}/2$,$0$,$\Delta_z$) and  $\mathbf{b}_4$=($0$,$a_{20}/2$,$-\delta z -\Delta z$), with $\delta z=0.18$ \AA, and $\Delta_z=1.08$ \AA, respectively. This way, $d_{10}=2.28$ \AA{} and $d_{20}=2.23$ \AA. Structure $B$ is obtained by swapping $x-$ and $y-$coordinates, making structures $A$ and $B$ degenerate \cite{Mehboudi2016,Barraza2018,Review} and the SnO monolayer ferroelastic \cite{Seixas2016}.

Point $C$ in Fig.~\ref{fig:figure1}(a) has lattice constants $a_1=a_2=a_C=(a_{10}+a_{20})/2$, and it sits midway through the path that joins points $A$ and $B$ (black dashed line). In structure $C$, O atoms are located at $\mathbf{b}_1$=(0,0,0) and $\mathbf{b}_2$=($a_C/2$,$a_C/2$,0), and Sn atoms are at $\mathbf{b}_3$=($a_C/2$,$0$,$\Delta_C$) and  $\mathbf{b}_4$=($0$,$a_{20}/2$,$-\Delta_C$) with $\Delta_C=1.17$ \AA, for an interatomic distance $d_C$=2.25 \AA. Figure \ref{fig:figure1}(b) shows side and top views of rectangular structures $A$ and $B$, and the (litharge) structure $C$.

Defining length variables $X$ and $Y$ as linear combinations of lattice parameters $a_1$ and $a_2$:
\begin{equation}\label{eq:transform}
X(a_1,a_2)=\frac{a_1-a_2}{\sqrt{2}}; \text{ } Y(a_1,a_2)=\frac{a_1+a_2-2a_C}{\sqrt{2}},
\end{equation}
one creates the energy cuts shown in Fig.~\ref{fig:figure1}(c) on the energy landscape along the dashed black line at the $(X,0)-$direction, and  the dashed red lines at the $(0,Y)-$ and $(\pm X_0,Y)-$directions indicated in Fig.~\ref{fig:figure1}(a).

The dependence of the landscape on $Y$ was almost identical along the red diagonal lines in Fig.~\ref{fig:figure1}(a) cutting through points $A$, $B$ and $C$: it is $J+c(X=0)Y^2$ around point $C$, and $c(X=X_0)Y^2$ around point $A$ (or $B$) with $c(X=0)$=23,042 K\AA${}^{-2}$/u.c. and $c(X=X_0)$=22,364 K\AA${}^{-2}$/u.c., for a negligible dependency of the coefficient $c$ on $X$; $c$ is approximated by  $(c(0)+c(X_0))/2$, to simplify the analytic description of the 2D energy landscape:
\begin{equation}\label{eq:profile}
U(X,Y)=aX^4 - bX^2 + cY^2 + \frac{b^2}{4a},
\end{equation}
with $a= 4,252$ K/\AA$^4$, $b=387$ K/\AA$^2$, and $c=22,703$ K/\AA$^2$ per unit cell.

Setting $U(X_0,0)=0$, $X_0=\pm\sqrt{\frac{b}{2a}}=0.21$ \AA, and inverting Eq.~\ref{eq:transform} at the analytic value of $X_0=0.21$ \AA, $Y=0$ \AA, one gets $a_{10}=3.99$ and $a_{20}=3.70$ \AA, which agrees with DFT results in Fig.~\ref{fig:figure1} reasonably well. $J\equiv{}U(0,0)-U(X_0,0)=\frac{b^2}{4a}=8.8$ K/u.c.

When expressed in units of K/u.c. $J$ turns out to be almost equal to $T_c$, as shown via {\em ab initio} MD performed at ambient pressure and at 12 predefined temperatures. { These calculations, performed using the NVT ensemble ({\em i.e.}, with a constant number of electrons and target pressure and temperature), require having two walls parallel to (not in contact with) the 2D material fixed throughout the MD evolution.}

\begin{figure}[tb]
\begin{center}
\includegraphics[width=0.48\textwidth]{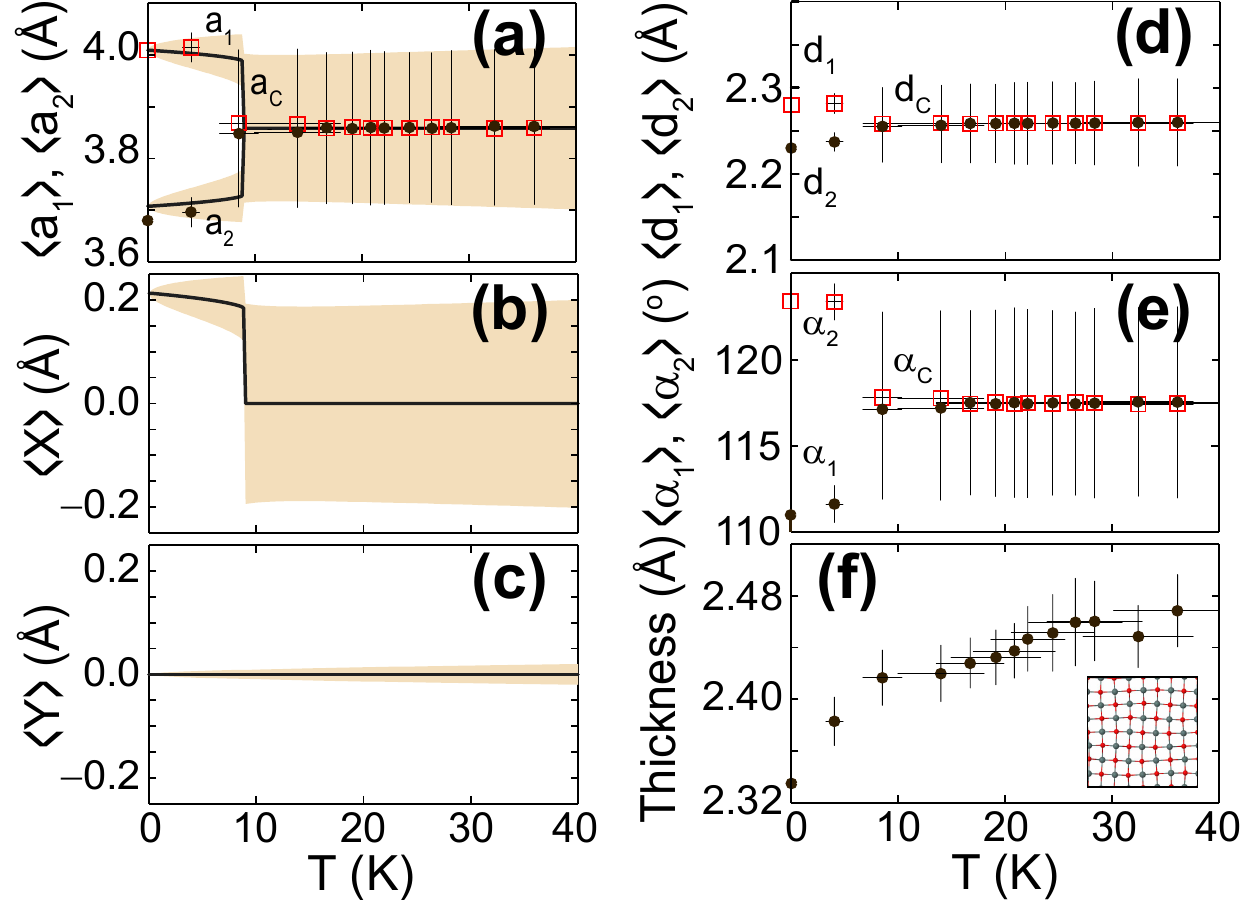}
\caption{(a) Thermal evolution of lattice constants from MD (data) and our model (solid trend lines). $T_c=8.5\pm 1.8$ K. (b) and (c) mean values and expectation values of $X$ and $Y$. (d-e) Thermal evolution of interatomic distances and angles defined in Fig.~\ref{fig:figure1}(b). (f) Thickness dependence on $T$; the inset is a structural snapshot during the thermal evolution.}\label{fig:figure2}
\end{center}
\end{figure}

\begin{figure}[tb]
\begin{center}
\includegraphics[width=0.48\textwidth]{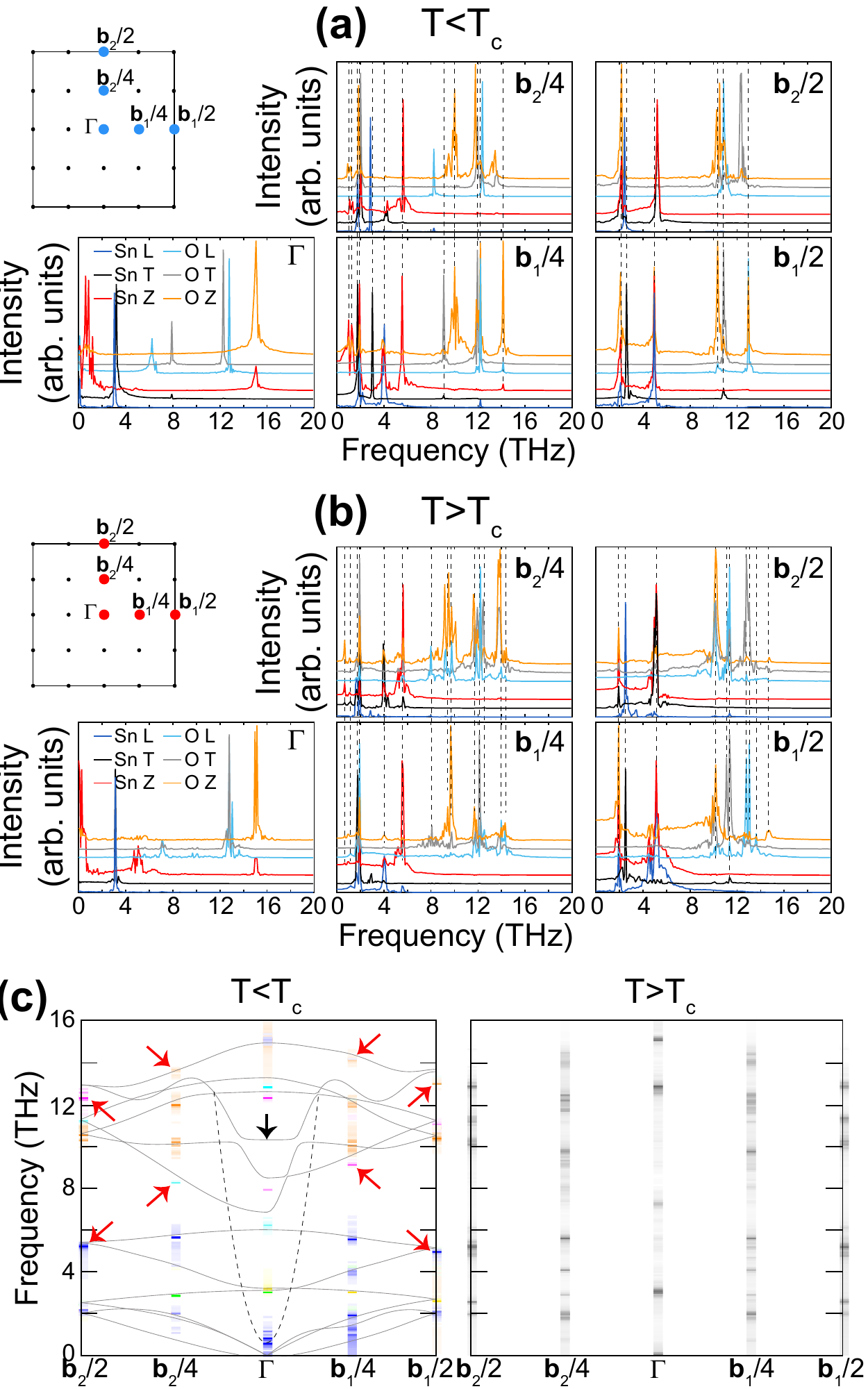}
\caption{PSD resolved over chemical element and spatial direction for (a) $T<T_c$ and (b) $T>T_c$. (c) Phonon dispersion for $T<T_c$ (left, mode resolved) and $T>T_c$ (right). The continuous trend was obtained {at $T=0$} using standard methods and is a guide to the eye. At least one phonon line above 10 THz at the $\Gamma-$ point on the left subplot (vertical arrow) appears to have gone soft to accumulate near zero frequency. Diagonal arrows showcase lack of symmetry when $T<T_c$.}\label{fig:figure3}
\end{center}
\end{figure}

{The {\em Qbox} MD code \cite{Qbox1,QboxPseudos} implements the plane-wave pseudopotential approach within DFT with highly-tuned PBE exchange-correlation pseudopotentials that have tested against multiple other plane wave codes for numerical consistency \cite{QboxPseudos}. It permits fixing the two walls not in direct contact with the SnO monolayer. MD data for charge-neutral SnO monolayers is presented in Figs.~\ref{fig:figure2} and \ref{fig:figure3} and discussed next.}

$T_c$ is signaled by a sudden change of lattice parameters $a_1$ and $a_2$ onto $a_C$, highlighting the importance of performing MD on an NPT ensemble to capture the transition temperature ($T$) \cite{Mehboudi2016,Mehboudi2016b,Barraza2018}.  MD calculations were performed on a $4\times 4$ SnO supercell originally set at configuration $A$, which assumes the existence of a macroscopic domain with such atomistic configuration (condition (iii) in Ref.~\cite{Review}). Using a 1.54 fs time resolution, we retrieved 16,000 frames, and used the last $2^{13}$ (=8192) frames to obtain ensemble averages and power densities in the frequency domain for any given temperature ($T$).

Figures \ref{fig:figure2}(a), \ref{fig:figure2}(d) and \ref{fig:figure2}(e) show the thermal evolution of lattice parameters ($\langle a_1\rangle$, $\langle a_2\rangle$), interatomic distances ($\langle d_1\rangle$, $\langle d_2\rangle$), and angles ($\langle \alpha_1\rangle$, $\langle \alpha_2\rangle$) (c.f., Fig.~\ref{fig:figure1}(b)), where $\langle\rangle$ indicates ensemble averages. There is a sudden collapse of these order parameters at $T_c=8.5\pm 1.8$ K, which ratifies the ansatz $T_c=\beta J$, with $\beta\simeq 1$ suggested early on. The thickness in Fig.~\ref{fig:figure2}(f) is the height difference among the lowest and the highest atom in the supercell, and it captures out-of-plane undulations \cite{Barraza2018}.

Together with earlier NPT MD results \cite{Mehboudi2016,Mehboudi2016b}, the value of $T_c$ obtained here points to the central role of the energy scale $J$ in understanding multiferroicity in 2D materials. The entirety of our MD data points to a relation $T_c \simeq J$.

It is illustrative to describe the structural transition right from the energy landscape in Fig.~\ref{fig:figure1}. Consider a particle there, and assign a mean kinetic energy equal to $k_BT/2$ to each of the two degrees of freedom ($k_B$ is Boltzmann's constant). In this classical construct, the particle has zero potential energy and a mean kinetic energy of $k_BT$ at point ($X_0,0$). The particle reaches the largest values of $X$ and $Y$ when its mean kinetic energy is zero, and the condition $k_BT=U(X,Y)$ determines the contours in the $(X,Y)$ plane the particle is confined to, as a function of $T$. Depending on whether $k_BT$ is smaller or larger than $U(0,0)=J$, the particle is either confined to a single well (monodomain $A$), or it oscillates among the two wells (i.e., the material has transitioned).

Equation $k_BT=U(X,Y)$ has four roots:
$X_{P\pm}(T)=\sqrt{(b\pm\sqrt{4ak_BT})/2a}$, and
$X_{N\pm}(T)=-X_{P\pm}(T)$,
where $N$ ($P$) stands for negative (positive). Being a classical construct, the elastic energy profile forbids direct tunneling among the two wells, so one is constrained to ($X_{min}(T)=X_{P-}(T)\le X\le X_{P+}(T)=X_{Max}(T)$) when $k_BT\le J$ at monodomain $A$. Both wells are accessible when $T>J$, and $X_{min}(T)=X_{N+}(T)\le X\le X_{P+}(T)=X_{Max}(T)$. $X_{min}(T)$ takes on two different values, depending on whether $T\le J$ or $T> J$.

Contours are given by:
$Y_{\pm}(X,T)=\pm\sqrt{(k_BT-aX^4+bX^2-J)/c}$, and ensemble averages within the model for any function $f(X,Y)$ are obtained from \cite{Kittel}:
\begin{equation*}
\langle f(X,Y)\rangle\equiv\frac{\int_{X_{min}(T)}^{X_{Max}(T)} \int_{Y_-(X,T)}^{Y_+(X,T)}e^{-U/k_BT} f(X,Y) dXdY}{\int_{X_{min}(T)}^{X_{Max}(T)} \int_{Y_-(X,T)}^{Y_+(X,T)} e^{-U/k_BT} dXdY}.
\end{equation*}
The solid lines in Fig.~\ref{fig:figure2}(a) are $\langle a_1\rangle$ and $\langle a_2\rangle$; shaded areas are standard deviations. These values were obtained from Figs.~\ref{fig:figure2}(b) and \ref{fig:figure2}(c) through an inversion of Eqn.~\ref{eq:transform}. The standard deviations on $a_1$ and $a_2$ are equal to 0.048 \AA, right below $T_c$. The agreement between the model and MD --including fluctuations-- is remarkable.

Despite of the small supercell size (inset in Fig.~\ref{fig:figure2}(f)), the power spectral density (PSD) --the Fourier transform of the MD data onto reciprocal space and frequency \cite{power}-- shown in Figs.~\ref{fig:figure3}(a,b) displays sharp peaks that uncover the phonon frequencies of the SnO monolayer, and include anharmonic contributions by construction. The PSD is resolved over chemical species (with slow modes mostly due to Sn, and overall faster modes due to O, which was vertically shifted in these plots) and orthogonal displacements $L$, $T$, and $Z$ \cite{Kittel}. The supercell provides a coarse sampling of reciprocal space that includes $k-$points $i\mathbf{b}_1/4+j\mathbf{b}_2/4$, where $\mathbf{b}_1$ and $\mathbf{b}_2$ are reciprocal lattice vectors and $i,j$ range in between $-2$ and $2$.

\begin{figure*}[tb]
\begin{center}
\includegraphics[width=0.96\textwidth]{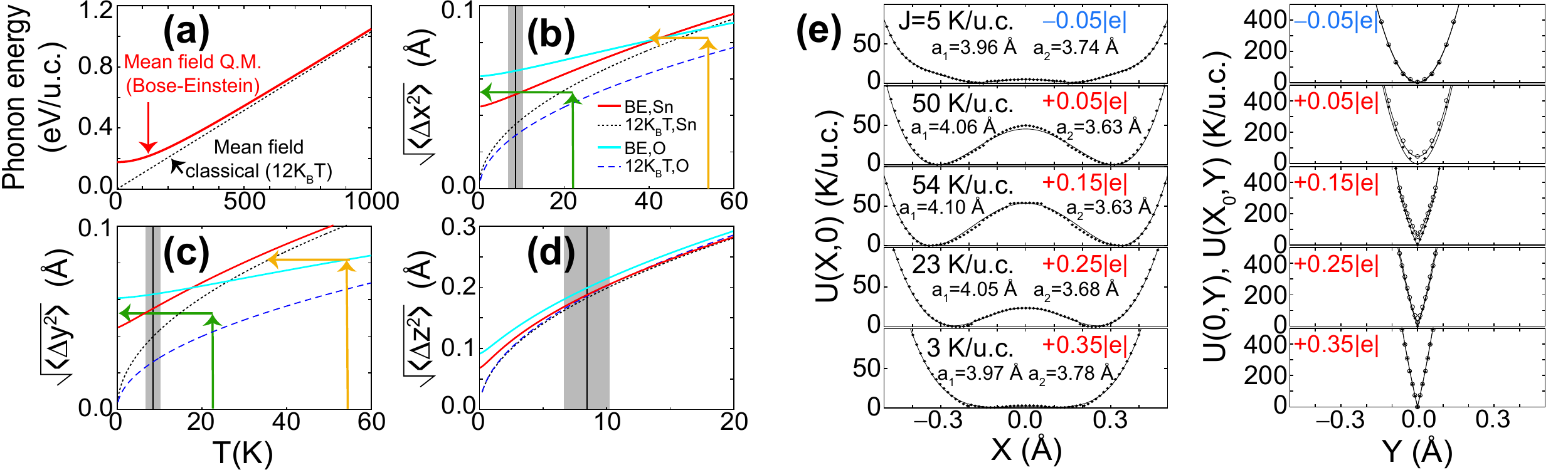}
\caption{(a) Energy due to phonons according to Bose-Einstein statistics, and to the classical occupation linear in $T$. (b-d) Atomic displacements of Sn and O atoms, calculated with Bosonic occupations (solid lines) and classical occupations. (e) Cuts {$U(X,0)$, $U(X_0,Y)$ and $U(0,Y)$ show the evolution of the elastic energy landscape and} the tuning of $J$ with charge doping.}\label{fig:figure4}
\end{center}
\end{figure*}

The rectangular unit cell imposes an asymmetry on the PSD shown in Fig.~\ref{fig:figure3}(a) at $T=4.0\pm 0.8$ K $< T_c$, that is best brought about by placing the spectral density at point $\mathbf{b}_2/4$ over that acquired from point $\mathbf{b}_1/4$, and that of $\mathbf{b}_2/2$ over $\mathbf{b}_1/2$, and aided by vertical dashed lines originating from plots $\mathbf{b}_1/4$ and $\mathbf{b}_1/2$. There, the Z-mode in orange at 14 THz at $\mathbf{b}_1/4$ appears at a lower frequency at $\mathbf{b}_2/4$. In a similar manner, the O $T$ mode seen in gray at about 8.5 THz at $\mathbf{b}_1/4$ is downshifted onto the $L$ mode in light blue at $\mathbf{b}_2/4$. A similar downshift is seen at the maximum frequency mode at $\mathbf{b}_2/2$ with respect to its value at $\mathbf{b}_1/2$. There is a significant spectral density at frequencies $\gtrsim$ 0 THz and at least two closely spaced peaks at the $\Gamma-$point in Fig.~\ref{fig:figure3}(a). As showcased at $T=16.7\pm 3.2$ K in Fig.~\ref{fig:figure3}(b), the PSD shows aligned peaks for $T>T_c$, thus reflecting an enhanced symmetry.

The power spectrum is rearranged into a phonon dispersion in Fig.~\ref{fig:figure3}(c) for $T<T_c$ (left, mode resolved) and $T>T_c$ (right), while the overlaid continuous trend is a phonon dispersion at 0 K, without anharmonic contributions, that was obtained from the Hessian matrix at $T=0$. Direct comparison of the continuous trend lines and the PSD data shows at least one missing phonon line above 10 THz at the $\Gamma-$ point on the left subplot (highlighted by a vertical arrow), which appears to have gone soft to accumulate near 1 THz frequency (as suggested by a dashed overlaid curve). The possibility of the mode at 10 THz being absent due to the small supercell employed still does not answer the existence of the large spectral density near 1 THz at the $\Gamma-$point seen in Figs.~\ref{fig:figure3}(a) and \ref{fig:figure3}(c), that does not match any phonon line suggested by the Hessian. Diagonal arrows highlight the reduced symmetry when $T<T_c$.

Ferroic behavior necessitates the feasibility of creating macroscopic monodomains \cite{Review}; i.e., that the two wells
on the elastic energy landscape are separated by a classical wall that forbids quantum tunneling. This classical limit is not guaranteed here due to the small value of $J$ and the presence of a light element (oxygen) prone to significant zero-point, quantum fluctuations.

Using the phonon spectrum and eigenvectors obtained from the Hessian, one can contrast the phonon energy and the mean displacements obtained using occupations consistent with Bosonic quantum statistics, with the occupations resulting from a classical occupation that is linear in temperature, and which underlies all the results presented up to this point. The drastic enhancement in occupation {when Bosonic statistics are considered} may drive a quantum paraelastic phase reminiscent of the quantum paraelectric phase in SrTiO$_3$ \cite{r1,r2,r3}.

The energy due to phonons with a Bosonic occupation (BE) shown by a red solid line in Fig.~\ref{fig:figure4}(a) displays the usual uptick at zero temperature; the classical occupation is linear in temperature, and it agrees with the quantum occupation {only} at large $T$. The Bosonic occupation in turn produces the uptick on the rms atomic displacements at low $T$ \cite{masri} depicted in Figs.~\ref{fig:figure4}(b-d), when compared with results obtained using occupations linear in $T$. The solid lines coalesce onto the dashed lines at large $T$. The solid vertical line in Figs.~\ref{fig:figure4}(b-d) is $T_c$, while the dashed area indicates its standard deviation.

Figure \ref{fig:figure4}(d) indicates that out-of-plane displacements $\langle\Delta z^2\rangle$ are not too dissimilar from both statistics. But in-plane components $\langle\Delta x^2\rangle$ and $\langle\Delta y^2\rangle$ obtained from the BE occupation show large excursions at $T=0$ already, only attainable with a classical occupation at higher $T$. Figure \ref{fig:figure4}(e) suggests that $J$ can be increased by {a slight} hole doping \cite{Seixas2016} and further decreased by {either} electron doping { or a substantial hole doping}: charge doping is a handle to get the SnO monolayer in and out of the quantum paraelastic phase.

\begin{figure}[tb]
\begin{center}
\includegraphics[width=0.48\textwidth]{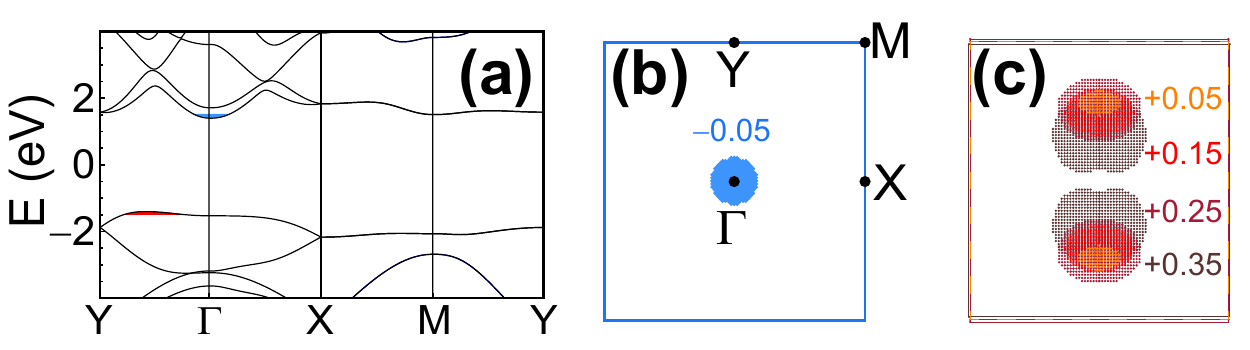}
\caption{(a) Electronic band structure of neutral SnO monolayer. Blue (red) shadowed regions point to locations of reciprocal space from which electrons are added (removed). (b) Electron doping occurs around the $\Gamma-$point but (c) hole doping first happens along a shallow pocket along the $\Gamma-Y$ line. Units in (b) and (c) are $|e|$/u.c. The Brillouin zone boundaries in (c) reflect the change of $a_1$ and $a_2$ upon doping.}\label{fig:figure5}
\end{center}
\end{figure}

{ $J$ depends on whether the charge added is symmetric or asymmetric on the $p_x$ and $p_y$ orbitals; c.f. Table~\ref{ta:t1}. The apparent correlation among $a_1/a_2$ and $J$ in Table \ref{ta:t1} is reminiscent of that given in Ref.~\cite{Mehboudi2016}. In turn, the location in reciprocal space where electron charge is added or removed seen in Fig.~\ref{fig:figure5} determines the orbital character.}

In the right side of Figs.~\ref{fig:figure4}(b,c), the in-plane rms displacement with Boson occupations may lead to a lower $T_c$, as suggested by the horizontal orange arrows in Figs.~\ref{fig:figure4}(b-c). The renormalization of $T_c$ is increased as $J$ is brought to smaller values, as displayed by larger horizontal green lines in Figs.~\ref{fig:figure4}(b-c). For values of $J$ smaller than 21 K/u.c., fluctuations of the order of 0.048 \AA --that drove the classical transition in Fig.~\ref{fig:figure2}(a)-- will occur at 0 K already, to set the quantum paraelastic phase.

\begin{table}
 \caption{Orbital character of added (negative sign) or removed (positive sign) electronic charge per unit cell, ratio among lattice parameters $a_1/a_2$, and  $J$ ($K$/u.c.).}\label{ta:t1}
\centering
\begin{tabular}{c|cccc|c|c}
\hline
$q$ ($|e|$) & $\Delta s$ & $\Delta p_x$ & $\Delta p_y$ & $\Delta p_z$   & $a_1/a_2$ & $J$\\
\hline
$-0.05$     & $-$0.019 & $-$0.001 & $-$0.001 & $-$0.029 & 1.06 & 5 \\
$0.05$     & 0.013 & 0.000 & 0.003 & 0.034 & 1.12& 50 \\
$0.15$     & 0.041 & 0.000 & 0.010 & 0.099 & 1.13& 54 \\
$0.25$     & 0.068 & 0.001 & 0.013 & 0.168 & 1.10& 23 \\
$0.35$     & 0.095 & 0.002 & 0.010 & 0.242 & 1.05& 3 \\
\hline
 \end{tabular}
\end{table}

There are experimental and engineering challenges related to sample quality and exfoliation down to monolayers that must be met to observe these phenomena. The SnO monolayer limit is yet to be achieved, though encouraging results demonstrating exfoliation down to 4 monolayers and 2 monolayers have been recently reported \cite{exp1,exp2}. SnO samples that show ambipolar doping \cite{exp3,vdW} will be ideal to test the present predictions, and additional control of electron doping, necessary to further explore the paraelastic phase, may be achieved via electrostatic doping with gates (e.g., Ref.~\cite{exp4}).

In conclusion, we have identified charge-neutral SnO monolayers as incipient quantum paraelastic materials; i.e., materials in which a ferroic structural configuration may be forbidden due to quantum fluctuations. It has been shown that the elastic energy barrier, expressed in K/u.c., is a reasonable estimate for the transition temperature of charge-neutral SnO monolayers. The power spectrum obtained directly from the MD evolution show missing modes, that go to lower energies prior to $T_c$ and herald the structural transition. The magnitude of $T_c$ may be tuned via charge doping, for quantum statistics to dictate the ferroic behavior of these materials.

T.B. was funded by the NSF (Grant No. DMR-1610126), A.P.S. by FONDECYT, project No 1171600 (Chile), P.D. by Laboratory Directed Research and Development (LDRD) funding from Argonne National Laboratory, provided by the Director, Office of Science, of the U.S. DOE under contract DE-AC02-06CH11357, and S.B.L. by the U.S. DOE, Office of Basic Energy Sciences (Early Career Award DE-SC0016139). Part of this work was performed at the Center for Nanoscale Materials, a U.S. Department of Energy Office of Science User Facility, and supported by the U.S. DOE, Office of Science, under Contract No. DE-AC02-06CH11357. Calculations were performed on Cori at NERSC, a U.S. DOE Office of Science User Facility operated under Contract No. DE-AC02-05CH11231. Conversations with P. Kumar, B. Fregoso and G. Naumis are gratefully acknowledged.


%

\end{document}